\documentclass[12pt]{spieman}  % 12pt font required by SPIE;
\usepackage{amsmath,amsfonts,amssymb}
\usepackage{graphicx}
\usepackage{setspace}
\usepackage{tocloft}
\usepackage{lineno}
\linenumbers
\title{The Case for High Resolution Spectroscopy in the Ultraviolet}

\author[a]{Jeffrey L. Linsky}
\author[b]{Todd M. Tripp}
\author[c]{Seth Redfield}
\author[d]{Kevin France}
\affil[a]{University of Colorado, JILA and Department of Astrophysical and Planetary Sciences, Campus Box 440, Boulder, Colorado, United States}
\affil[b]{University of Massachusetts, Department of Astronomy, Amherst, Massachusetts, United States}
\affil[c]{Wesleyan University, Astronomy Department and Van Vleck Observatory, Middletown, Connecticut, United States}
\affil[d]{University of Colorado, Laboratory for Atmospheric and Space Physics, Boulder, Colorado, United States}

\cftpagenumbersoff{figure}
\cftpagenumbersoff{table} 
\begin{document} 
\maketitle
\nolinenumbers

\begin{abstract}
The Astro2020 Decadal Survey declared that the baryon cycle is one of the top-priority science topics for current astrophysics. Space instruments with both high spectral resolution and high throughput in the ultraviolet are required for investigations of low density warm and cold gas present in both the \underline{inner} regions of the baryon cycle (interstellar medium, star-exoplanet interactions, pre-main sequence stars, stellar winds, flows and structures driven by supernovae) and the \underline{outer} regions (outflows of matter and energy from galaxies, circumgalactic media). The Space Telescope Imaging Spectrograph on HST has pioneered such studies, but STIS has its limitations and the lifetime of HST is limited. There is a pressing need for future large instruments with high spectral resolution ($R\approx 100,000$) in the 120--320~nm wavelength band such as the present STIS E140H and E230H capabilities, but with increased throughput to study gas in sight lines to faint sources such as M dwarf stars and circumgalactic clouds. Multi-object spectroscopy at high spectral resolution could enhance observational efficiency. This document describes some of the scientific results obtained with STIS and the new science that an enhanced instrument on a large telescope such as HWO could accomplish. We provide examples of the resolution needed for these science investigations.
\end{abstract}

% Include a list of up to six keywords after the abstract

% Include email contact information for corresponding author
{\noindent \footnotesize\textbf{*}Jeffrey L. Linsky,  \linkable{jlinsky@jila.colorado.edu} }

\begin{spacing}{2}   % use double spacing for rest of manuscript

\section{Introduction}

High-resolution spectroscopy in the ultraviolet can be a finely crafted tool for understanding critical issues in astrophysics. While not an edge-of-the-universe tool (like photometry and low-resolution spectroscopy are for observing very faint sources), high-resolution spectroscopy can study low speed gas flows, shocks, and inhomogeneous physical properties that are beyond the capabilities of observations at lower spectral resolution. In particular, high resolution is needed to separate fine structure transitions (useful for electron density measurements), isotopic shifts (i.e. $^6$Li from $^7$Li), vibration-rotation lines of diatomic molecules without hydrogen  (e.g., CO), and both cold and warm gas with closely separated narrow absorption lines. High-resolution spectra can also be used to 'create a picture' of the three-dimensional structure of gas that is too faint to image directly (e.g., gas in the local ISM) or below the angular resolution scale of astronomical images (e.g., in the inner regions of proto-planetary disks)

An insidious problem occurs when narrow lines are blended and close to saturation; lower resolution spectra suffer from large errors in estimates of the column density and kinematic structure of the absorbing medium. At any resolution, a robust characterization of the line spread function is critical when fitting blended lines near saturation.

The ultraviolet spectral region consisting of the near-UV (roughly 170--320~nm), far-UV (120--170~nm), and Lyman-UV
(91.2--122~nm) contains a wide variety of spectral lines produced in thermal plasmas at temperatures of 3,000~K to $10^6$~K. For example, transitions from the ground states of abundant neutrals (H~I, D~I, C~I, N~I, O~I) and important ions including C~II, C~III, C~IV, N~V, and O~VI are present in the UV. Absorption lines from ground states provide direct measurements of column densities and thus relative abundances without models required for analyzing absorption transitions from excited states. Also, important molecules such as H$_2$ and CO have electronic transitions in the ultraviolet.

Combining the capabilities of high spectral resolution and UV spectral range opens the door to understanding important scientific questions as described below, but there are challenges. Very efficient optical coatings are essential, because high-resolution spectrographs typically have many reflections and optical coatings in the UV can have only modest reflectivity, especially at shorter wavelengths. The Cosmic Origins Spectrograph (COS) on HST is an example of a very efficient UV spectrograph with only one optical surface, but its resolution is modest (R=$\lambda/\Delta\lambda=18,000$). Since high-resolution spectra disperse light over a large number of detector pixels, all astronomical targets are faint and therefore require telescopes with large diameter primary mirrors. Cube-sats and small telescopes are inadequate. The present instrument that meets these requirements is the Space Telescope Imaging Spectrograph (STIS) installed on HST in 1997 with the highest $R=114,000$ in the near-UV and far-UV. After 28 years it is still operating well, which is a strength of UV spectrographs --- the detectors and optics can be built to last for a long time. But how long HST will be operating in the future is unknown. Now is the time to decide how high-resolution UV spectroscopy will be accomplished in the post-STIS era whether on HWO or another large mission. In particular, what will be its enhanced capabilities in spectral range and throughput to accomplish critical astrophysics objectives.

Since gas flows often are cold or have low speeds, a resolution of $R\approx 100,000$ (3 km~s$^{-1}$) is highly desirable. The need for spectral resolution this high is demonstrated by the observed spectrum of the star $\kappa^1$~Cet at the top of Figure 1, which shows interstellar Mg~II absorption at three velocities separated by 13 km~s$^{-1}$ [1]. This STIS spectrum obtained with the E230H grating ($R=114,000$) at a velocity resolution of 2.63 km~s$^{-1}$ cleanly separates the three components, permitting individual measurements of each of these velocity components, but lower resolution spectra would blend the velocity components and make it difficult to measure line widths and even determine whether there is more than one velocity component. 

The spectrum shown at the bottom of Figure 1 is the observed spectrum smoothed to resolution $R=30,000$. While the absorption is notably asymmetric, it nonetheless is best fit by a single component. The modest improvements in $\chi^2$ for a 2 or 3 component fit at this low resolution are not significant based on an F-test. Because of this, the $R=30,000$ resolution spectrum clearly misses the weakest 7 km~s$^{-1}$ feature and it splits the difference between the 13 km~s$^{-1}$ (Hyades cloud) and 21 km~s$^{-1}$ (Local Interstellar Cloud) features with a single component at 18 km~s$^{-1}$. 

The low resolution emission line reconstruction also suffers because the self-reversal in the MgII emission line is less obvious. Therefore, the intrinsic stellar flux is overestimated in the $R=30,000$ spectrum. Finally, the total column density (see Table~1) is significantly overestimated, because the absorption is close to being optically thick. The $R=114,000$ spectrum clearly indicates that the interstellar absorption lines are optically thin, and that the total column density is log$_{10}$ [N(MgII)] =12.61 when all three components are combined. Due to the blending of the two features in the low-resolution spectrum, the fitting compensates by inferring a deeper absorption feature that is now optically thick and overestimates the column density by a factor of 70! {\bf This is a good demonstration of how blending and absorption near optical depth unity can wreak havoc on obtaining accurate physical measurements.}
 
The observed spectrum smoothed to a resolution $R=50,000$ is shown in the middle panel of Figure~1. Here we identify the two strong components, but do not detect the weaker feature at 7 km~s$^{-1}$. Perhaps the biggest problem is the overestimation of the column density at lower resolution when there are blended components near optical depth unity. Here the total column density is 13.99, a factor of 20 larger than the total column density inferred from the original high-resolution spectrum. Also, the line widths ($b$) are very much smaller than inferred from the original high-resolution spectrum, because of the degeneracy between column density and line width near optical depth unity.

\begin{figure}
\begin{center}
\begin{tabular}{c}
\includegraphics[height=12.0cm]{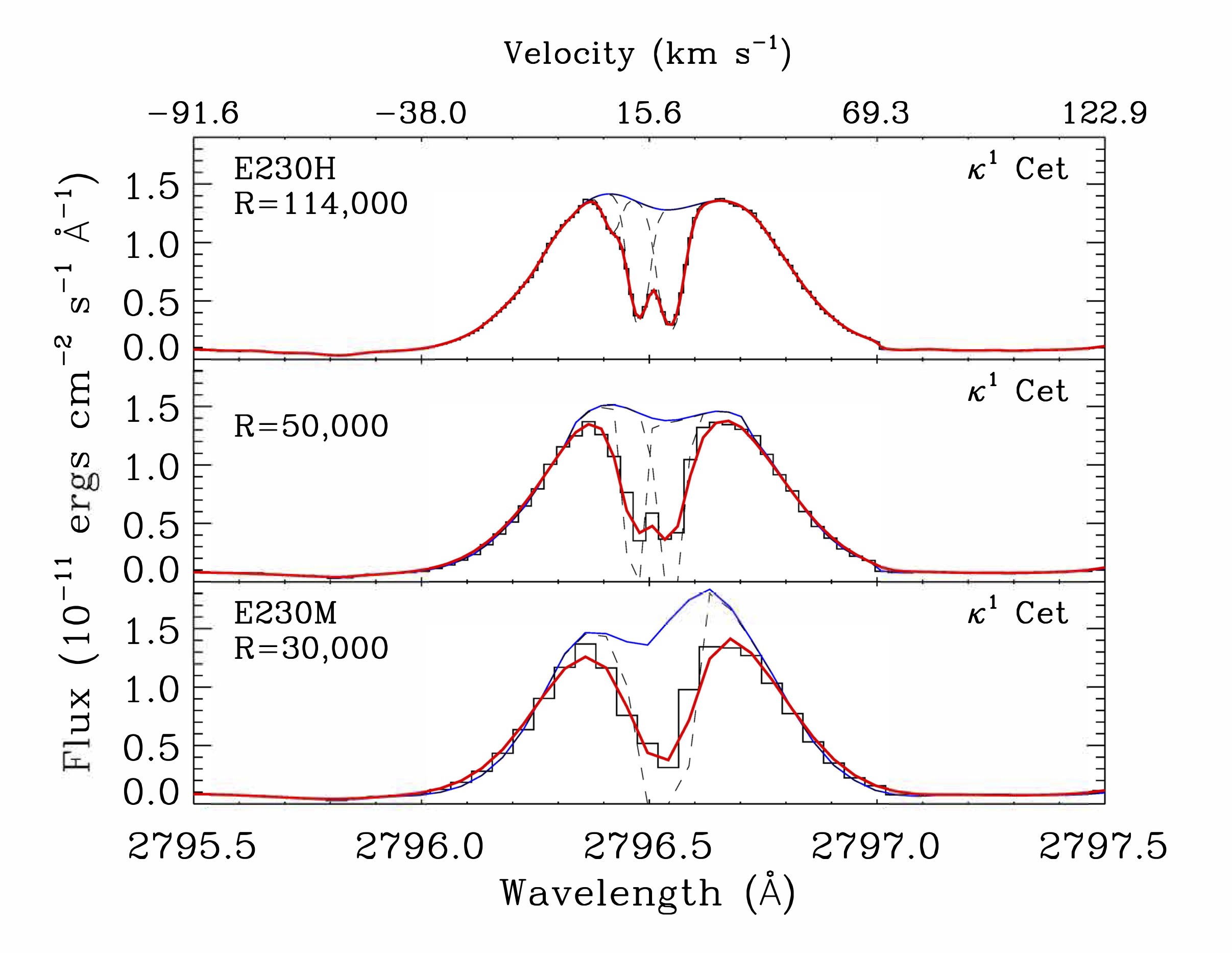}
\end{tabular}
\end{center}
\caption 
{\label{Resolution}
{\em Top panel:} Spectrum of the MgII line of $\kappa^1$~Ceti observed with high resolution (STIS E230H). {\em Middle panel:} Same spectrum degraded to 50,000 resolution. {\em Bottom panel:} Same spectrum degraded to 30,000 resolution. Dashed lines are for the inferred absorption lines before instrumental broadening, and the red lines are with instrumental broadening.} 
\end{figure}

 \begin{table}[ht]
\caption{Comparison of spectra at different spectral resolutions.} 
\label{Resolution}
\begin{center}       
\begin{tabular}{lllll} 
\hline
Resolution & Component & $v$ (km~s$^{-1}$) & $b$ (km~s$^{-1}$) & log column density\\
\hline\hline
100,000 & 1 & 20.78 & 3.26 & 12.34\\
100,000 & 2 & 13.33 & 2.68 & 12.21\\
100,000 & 3 & 7.25 & 2.48 & 11.42\\ \hline
50,000 & 1 & 20.71 & 2.21 & 13.15\\
50,000 & 2 & 12.67 & 1.07 & 13.92\\ \hline
30,000 & 1 & 18.2 & 3.0 & 14.47\\
\hline 
\end{tabular}
\end{center}
\end{table} 

We list below some of the major results obtained from the analysis of high-resolution UV spectra. This is far from a complete list, but it is representative of the science that can now be obtained with this capability. 

\section{Interstellar medium}

Many interstellar medium issues require high-resolution UV spectra because at very low densities the population of atoms and ions are entirely in their ground states, and absorption lines from these ground states are almost entirely in the UV. Absorption line spectroscopy allows the identification of trace interstellar species against the bright UV continuum or emission lines of target stars. Some research highlights of interstellar medium:

\subsection{Kinematics}

 The local region of the Galaxy, often called the Local Interstellar Medium (LISM), includes the partially ionized gas extending outwards from the Sun for 5--10~pc that is embedded inside a supernova bubble called the Local Bubble with fully ionized hydrogen, which is in turn is surrounded by a low ionization shell. The interstellar radial velocities inferred from high-resolution spectra of nearby warm gas are not random, but rather are consistent with velocity vectors of individual structures called clouds. Originally two clouds were identified kinematically: the Galactic or G cloud [2] in the direction of the Galactic Center, and the Local Interstellar Cloud (LIC) [3] mostly in the region opposite to the Galactic Center. Redfield \& Linsky [4] identified 15 velocity vectors of clouds within about 10 pc of the Sun, but there may be additional clouds that have not been identified because (a) there are too few velocity components inconsistent with the 15 velocity vectors and/or (b) the clouds are more distant covering a small angular patch of the sky making it difficult to fit a velocity vector from only radial velocity data. 
 
 {\bf Science Objective} Whether the multi-cloud model consisting of clouds each with its own co-moving gas or a single cloud model with internal velocity gradients proposed by Gry and Jenkins [5] is a better representation of the data remains an open question [6] that requires high-resolution spectra of many more sight lines to fainter stars including  M dwarfs. This requires a large aperture telescope feeding a high-resolution UV spectrograph to obtain S/N$\geq 20$ per resolution element at a flux level of $10^{-15}$ erg~cm$^{-2}$~s$^{-1}$~\AA$^{-1}$ in the Mg~II and Fe~II lines and ten times that flux level at the Lyman-$\alpha$ line.
 The relative importance of directed flows including shocks, magnetic fields, and ionizing radiation from different directions can be tested by observing the kinematic structure of the local warm gas.

\subsection{Morphology} 

The column density of neutral hydrogen along a sight line to a target star can be measured either from absorption in the deuterium Lyman-$\alpha$ line at --82 km~s$^{-1}$ from the center of the H~I Lyman-$\alpha$ line [7] or from metal absorption lines such as Mg~II and Fe~II [8]. Since the optical depth in the core of interstellar H~I Lyman-$\alpha$ line is always greater than $10^5$ even for the nearest target stars, the deuterium line with D/H$=1.56\times 10^{-5}$ [9] is an excellent proxy for the neutral H column density. Since the deuterium Lyman-$\alpha$ line can also be optically thick, spectra of the less opaque Lyman-$\beta$ (102.5nm), Lyman-$\gamma$ (97.25~nm), and higher lines in the Lyman ultraviolet region would yield very accurate hydrogen column densities.

Metal lines are less accurate proxies for H as depletions are highly variable in the LISM. Although moderate spectral resolution (6 km~s$^{-1}$) is adequate to measure the deuterium line profile and separate it from the broad hydrogen line, high-resolution (3 km~s$^{-1}$) spectra of metal lines like Mg~II and Fe~II are needed to identify multiple velocity components that are smeared out by the low mass deuterium. Otherwise deuterium column densities are inaccurate when the line is partially saturated.

Unfortunately H column densities along a sight line to a target star do not provide information on where a cloud begins or the extent of a cloud before and potentially beyond the distance to the star. 

{\bf Science Objective} While two-dimensional maps of the cloud structure in the first 10~pc from the Sun have been constructed with existing data and assumptions about the filling factors of neutral hydrogen in many sight lines, the real need is for three dimensional maps based on realistic assessments of filling factors. A critical tool for this is the determination whether or not neutral hydrogen gas surrounds the astrosphere of the star at the end of a sight line [10]. This requires the detection of red-shifted absorption in the hydrogen Lyman-$\alpha$ line in a hydrogen wall created by charge exchange between hydrogen atoms inflowing from the LISM and protons. The few detections of this feature demonstrates the feasibility of the technique, but high-resolution spectra of many more sight lines including towards faint M dwarfs are needed.

\subsection{Mean thermal properties}
\label{sec_mean_thermal}

Fitting interstellar absorption lines with Gaussian or Voigt profiles yield line widths $b$ that are composed of thermal and non-thermal components, $b^2=3kT/m+\xi$, where $T$ is the temperature, $m$ is the atomic mass, and $\xi$ is the turbulent velocity. The separation of $T$ from $\xi$ is accomplished by observing a line of a low mass specie (typically D~I Lyman-$\alpha$), which is dominated by thermal broadening, and a line or lines of high mass ion (Mg~II or Fe~II), which is dominated by subsonic turbulent broadening. Since the Lyman-$\alpha$ line is at 121.6~nm, the Fe~II lines are at 260~nm, and the Mg~II lines are at 280~nm, a wide spectral range in the ultraviolet is necessary for these measurements. The mean temperatures of the 15 clouds [4] range from 3,000~K to 10,000~K. Since the clouds are partially ionized and have a wide range of mean temperatures, they are not easily fit by classical models of warm neutral gas and warm ionized gas. 

Neutral hydrogen number densities cannot be measured directly, but can be inferred by considering likely filling factors, the ratio of neutral hydrogen column density $N$(HI) to distance to the star in sight lines. If the mean neutral hydrogen number density $<n({\rm H~I})>\approx 0.2$~cm$^{-3}$, as is inferred for the hydrogen flowing into the heliosphere, then except for the star AD~Leo all of the filling factors $f=N$(H~I))/(0.2d) are in the range $f\approx 0.5$ within about 4~pc of the Sun and decreasing out to 10~pc. Linsky et al. [10] argued that a more likely density in the clouds within 4~pc is $<n({\rm H~I})>\approx 0.1$~cm$^{-3}$. Strong support for this result is that there are four stars within 4~pc that show only LIC velocities and have detected hydrogen walls showing that they are surrounded by neutral hydrogen in the LIC. This implies that $f=1$ in these sight lines and therefore 
$<n({\rm H~I})>\approx 0.1$~cm$^{-3}$. This new value of $<n({\rm H~I})>$ means that the local clouds are twice as large as previously thought and thus will likely overlap in space.

{\bf Science Objective} The dominant heating processes in the LISM have been identified as collisions of cosmic rays with dust and gas and the radiation from dust grains following absorption of stellar radiation. An important question is whether another heating source may be present. Shull \& Kulkarni [11] showed that the bow shocks and wakes produced by stars moving rapidly through the LISM can both stir and heat the gas. Wakes called proplyds are seen for example in the Orion Nebula where fast moving young stars with large mass loss rates plow through the ISM, but such wakes have never been detected from the moderate age stars with small mass loss rates in the LISM. High spectral resolution and higher throughput than is possible with STIS are needed to address this question.

\subsection{Inhomogeneous properties}

Within a cloud, temperatures are inhomogeneous with some sight lines in the LIC as low as 3,000~K and some as high as 12,500~K. The spatial scale for significant temperature differences is less than 5,100 au [10]. With a Larmor radius of only 350 km = 0.0017 AU for a thermal proton in a $3\mu$G field, interstellar magnetic fields can maintain inhomogeneous plasmas for a considerable time.

{\bf Science Objective} Until now typical analyses of interstellar plasmas have assumed that the plasma is thermal and that the widths of interstellar lines consist of two components: a component proportional to $m^{-0.5}$ called temperature and a component independent of atomic mass called turbulent broadening. At the very low densities in the LISM, $n_H\approx 0.1$ cm$^{-3}$, the time scales for ionization and excitation equilibria can be much longer that the time scales for changes in local densities and photoionization rates.  Adiabatic cooling of electrons, which is much faster than the recombination rates, produces a plasma with ionization far from thermal equilibrium. Recombining plasmas following supernova explosions and shock waves can lead to plasmas that contain a supra-thermal component. Supra-thermal plasmas have never been measured in the LISM, but they could be with the right instrument. Supra-thermal particles produce broad absorption line wings compared to thermal plasmas. Measurement of the shape of these $\kappa$-shaped profiles would yield both the energy and abundance of the supra-thermal particles. High spectral resolution and high S/N are needed for this experiment. The pressure of supra-thermal particles may be important component of the total pressure in the LISM.  

{\bf Science Objective} Neutral hydrogen column densities $N$(HI) along some sight lines are factors of 2 to 6 times larger than other sight lines at similar distances. Swaczyna et al. [12] showed that high $N$(HI) along two sight lines can be explained by the overlap of the LIC and G clouds. High $N$(HI) along the sight lines to 6 additional targets [8] can also be explained by the LIC/G overlap region, but another 12 high $N$(HI) sight lines require another explanation, perhaps the overlap of other clouds, shocks, wakes of fast moving stars through the LISM [11], or shells produced by supernovae [13]. High resolution spectra of a large number of sight lines are needed to address this mystery.

\subsection{Cold gas in the ISM}

Jenkins \& Tripp [14] showed that the fine structure lines of neutral carbon are sensitive diagnostics of gas in the diffuse cold ($T\approx 80$~K) neutral component of the ISM. High-resolution STIS E140H spectra of fine structure CI lines reveal the distribution of thermal pressures that include a small component with pressures 100 times higher than most cold gas. The amount of the very high-pressure regions is correlated with the local ultraviolet starlight density, but high gas pressures
are also found in regions well away from stars [15]. Most of the cold gas is in small structures with supersonic turbulent motions, but the heating of the gas, possibly by the cascade of the turbulent motions into heat, and the confinement mechanism for the very high pressure component are uncertain.

\begin{figure*}
\includegraphics[width=14cm]{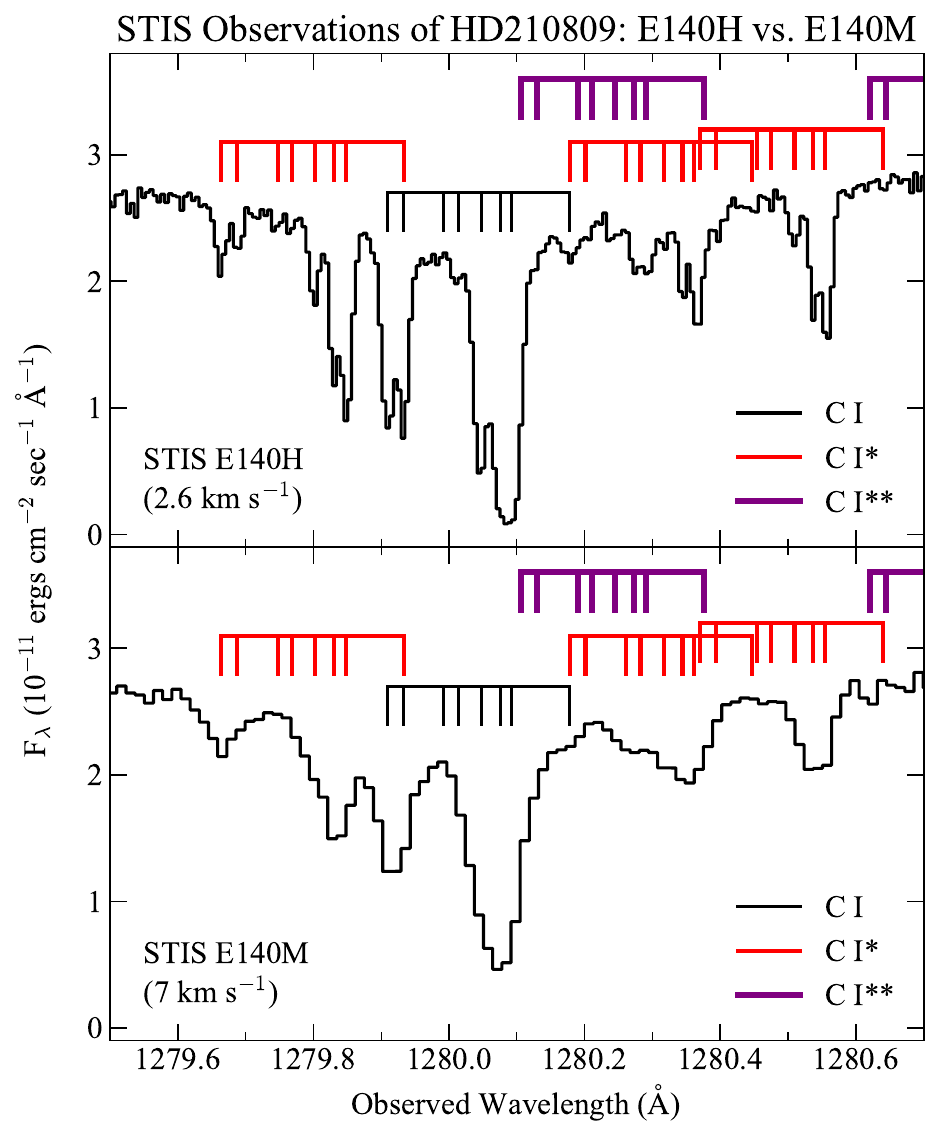}
\caption{Comparison of STIS E140H (upper panel) vs. STIS E140M (lower panel) observation of the C~I multiplet at 1280 \AA\ in the spectrum of HD210809.  The benefits of the higher spectral resolution of E140H are immediately apparent: in the E140H spectrum, discrete components have mostly well-constrained line shapes, and with the additional constraints from other C~I multiplets in the UV, the column densities of C~I, C~I*, and C~I** can be determined (Jenkins \& Tripp 2001).  Unfortunately, despite reasonably the good resolution of 7 km s$^{-1}$ in the E140M spectrum, crucial information is lost, and attempts to extract gas pressures from the lower-resolution spectrum suffer from serious systematic errors. \label{fig_CI}}
\end{figure*}

The measurement of absorption lines of C~I, C~I*, and C~I** (i.e., the absorption out of the three fine-structure levels of the neutral carbon ground state) is very challenging for several reasons.  The low temperatures of the gas lead to narrow line widths, and consequently excellent spectral resolution is required to accurately measure the line properties.  Moreover, the atomic physics of these transitions leads to complicated and overlapping absorption profiles (see, e.g., Figure~4 of [16]).  Fortunately, there are many C~I multiplets with a range of line strengths available in the UV, and since the relative locations of the C~I, C~I*, and C~I** lines change from one multiple to the next, Jenkins \& Tripp [16] were able to devise a method to reliably extract the column densities in the three fine-structure levels, but, this method also requires high spectral resolution to yield robust results.  To illustrate why very good resolution is required, Figure~\ref{fig_CI} compares observations of the interstellar C~I multiplet at 1280 \AA\ in the spectrum of the star HD210809, obtained with the STIS E140M (7 km s$^{-1}$ resolution) and STIS E140H (2.6 km s$^{-1}$) spectrographs.  Toward this star, eight components are detected in C~I absorption, which leads to a complex amalgamation of absorption from the three atomic levels.  However, with the 2.6 km s$^{-1}$ resolution of E140H (upper panel of Figure~\ref{fig_CI}), the various components appear as mostly discrete and well-constrained features. While there is considerable overlap of C~I, C~I*, and C~I** transitions, the Jenkins \& Tripp method can be successfully applied (including many other C~I multiples not shown in Figure~\ref{fig_CI}), leading in turn to measurements of the thermal gas pressure. On the other hand, in the 7 km s$^{-1}$ E140M spectrum (lower panel), much of the crucial information is lost, and it is not possible to sort out both the component structure and the C~I fine-structure excitation.  Studies of molecular absorption lines in the UV similarly benefit from excellent spectral resolution.

{\bf Science Objective} The much higher spatial resolution (less than 0.1 arcsec in the UV) of HWO compared to HST  should allow a UV spectrograph with 3 km~s$^{-1}$ resolution to study the small scale pressure and temperature fluctuations in warm and cold gas regions that should characterize the physical processes that confine and heat the gas. In particular, the ratios of neutral to ionized carbon and neutral to molecular hydrogen, all measurable with UV spectra, should provide tests of different proposed heating and pressure confinement mechanisms.  With greater collecting area than HST, HWO will also obtain observations of extragalactic absorption systems (e.g., damped Lyman $\alpha$ absorbers which often exhibit absorption lines of C~I and other neutral atoms), with quality and resolution similar to the upper panel in Figure~\ref{fig_CI}. Sub-arcsecond angular resolution coupled with high spectral resolution is needed to isolate stars in tightly packed regions such as globular clusters and star forming regions in distant galaxies. This capability in the near-IR could study molecular species including H$_2$ in star forming regions.

\section{Stellar Activity and Star-Exoplanet Interactions}

A critical question concerning whether exoplanets could be habitable is whether they can retain their atmospheres and thereby maintain liquid water on or just below their surfaces. There are several mechanisms for driving mass loss from exoplanet atmospheres including hydrodynamic escape resulting from the host star's extreme-UV radiation (e.g., [17,18]). Theoretical models of exoplanet mass loss now indicate that M star exoplanets in their habitable zones will not retain their initial atmospheres [19], but such models need to be tested since exoplanet magnetic fields may be important in retaining their atmospheric water [20]. Spectroscopy during transits is a powerful tool for identifying constituents in an exoplanet's atmosphere, since the emission from the host star supplies a bright background but variable stellar activity is a source of confusion [21]. 

High-resolution spectra of the Lyman and Werner bands of H$_2$, which is the dominant specie in young mini-Neptunes and Earth-mass planetary atmospheres are prominent in the UV [22]. Also, in the UV are the electronic bands near 95~nm of the homo-nuclear molecule N$_2$, which is the dominant constituent of the atmosphere of present day Earth and could be the dominant specie in habitable Earth-like exoplanets.

{\bf Science Objective} The modest resolution infrared spectra now being obtained by JWST are identifying constituents including NH$_3$, CO$_2$, and CO in exoplanet atmospheres during transits of warmer host stars. The search for molecules during transits of the much fainter M dwarfs is ongoing, but these spectra do not have the resolution needed to measure outflow velocities and thus mass-loss rates. High-resolution spectra of the hydrogen Lyman-$\alpha$ line and lines of MgII, CII, and other metals offer the possibility for measuring outflows predicted to have velocities of order 10 km~s$^{-1}$. Measurements of hydrogen velocities in the mass-loss tails of exoplanets observed in absorption after the end of transit would provide a clear measure of mass-loss rates.

{\bf Science Objective} Photo-chemistry in exoplanet atmospheres is driven by UV radiation from their host stars. In particular, the properties (spectral energy distribution and opacity) of hazes seen in M dwarf exoplanet atmospheres and predicted for Archean Earth are highly sensitive to the UV radiation of their host stars [23]. While Lyman-$\alpha$ and other bright emission lines are measured in the MUSCLES and Mega-MUSCLES surveys of M dwarfs with HST [24], the radiation likely present in the many weak UV lines and continuum is difficult to measure in these spectra given the noise at low flux levels. To obtain accurate measurements of the UV flux in the many weak lines and continuum of M stars with known and newly discovered exoplanets, there is a need for deep high-resolution spectra. Spectral resolution of 100,000 is needed to separate the many faint narrow lines from the continuum [25].

\section{Circumstellar Phenomena}

ALMA has observed sub-millimeter images of young stars with disks that often contain rings and open regions between the rings. Important phenomena include both accretion and mass loss in proto-stellar disks and perhaps the addition of molecules from comets in the debris disks of somewhat older stars. Rings in proto-stellar disks likely indicate planetary formation, which could be studied from flows observed in high-resolution spectra. Moderate resolution UV spectra of pre-main sequence stars are available from the HST archive, but there are few available spectra with 3 km~s$^{-1}$ resolution. Proposed accretion mechanisms leading to planetary formation are not yet confirmed by observations. Most of the mass in proto-stellar disks is H$_2$, and the Lyman and Werner bands of H$_2$ are present in the 91.2--160.0~nm spectral region.

{\bf Science Objective} The formation of planets in the disks of pre-main sequence stars could be tested from the likely low velocity flows and turbulence in the disks. Seen in absorption at modest resolution against the bright stellar Lyman-$\alpha$ line are fluorescent absorption lines of H$_2$ formed in the disks of the T~Tauri stars DF~Tau and V4046~Sgr [26]. High resolution is needed to identify the velocity structure in these absorption features produced by planets if present. High resolution is also needed to study density and temperature diagnostics of gas in these disks.

{\bf Science Objective} An important question is whether the composition of debris disks is first generation, mostly H$_2$ from the proto-stellar disk phase, or second generation, resulting from the evaporation of comets. The observed neutral carbon to molecular CO ratios (CI/CO) obtained from STIS E140H spectra of disks observed edge-on [27] are inconsistent with present models and leave the question of first or second generation composition open and the roles of stratification and mixing unanswered. High-resolution UV spectra of the electronic bands of H$_2$ and CO together with spectral lines of neutral hydrogen, carbon, and oxygen in the disks of both edge-on and other orientations could answer these questions. Also, high-resolution spectra of proto-planetary disks is required to understand the evolution of circumstellar gas. In the LUVOIR Report (2019 [28] such measurements were identified as one of the '100-hr highlight' science cases, and see also [29,30].

\section{Astrospheres}

The outer region of the heliosphere is often called the very local interstellar medium (VLISM) because the medium consists of gas flowing in from the LISM then modified by charge-exchange and other processes. Charge exchange reactions between interstellar neutral hydrogen and protons produces secondary neutral hydrogen atoms than are heated and slowed down with respect to the inflowing gas with enhanced density in a region called the hydrogen wall. Hydrogen Lyman-$\alpha$ absorption red-shifted from the main interstellar absorption feature has been detected in a few stars indicating that hydrogen walls are present.

{\bf Science Objective} Theoretical models suggest that the solar hydrogen wall in the heliosphere is located near $300 R_{\rm Sun}$, but Voyager~1, the most distant spacecraft, will not be operational when it reaches this distance. A good test of the location of the hydrogen wall would be imaging the Lyman-$\alpha$ line emitted by nearby stars. For $\alpha$~Cen~A, which is similar to the Sun at a distance of 1.3~pc, the angular diameter of the hydrogen wall should be about 1 arcsec compared to the 0.01 arcsecond resolution anticipated for HWO in the ultraviolet. An imaging high-resolution spectrograph on a large telescope would have the resolution to image many nearby stars to provide tests of astrosphere models. This type of observation could also test theoretical models of stars located in fully ionized interstellar and high density cold gas. 

\section{Circumgalactic Media}

The science applications of high-resolution UV spectroscopy presented above are examples of studies of the \underline{inner} regions of the so-called ``baryon cycle'' (the intricate cycle of gas inflows, star- and black-hole formation, and subsequent outflows of matter and energy) that regulates the evolution of galaxies and enables their complex star-formation histories.  The Astro2020 Decadal Survey declared that the baryon cycle is one of the top-priority science topics for current astrophysics, but many aspects of the cycle are difficult to study due to the low density and complex physics (including plasma physics) of the gas flows.  The \underline{outer} reaches of the baryon cycle --  the ``circumgalactic media'' (CGM) of galaxies -- are expected to have particularly low densities and physical conditions that make observations very challenging: like the ISM discussed above, the primary observational probes of the CGM require ultraviolet or X-ray observations from space.  Moreover, the CGM is known to consist of multiple gas phases ranging from hot phases ($T = 10^{5}$ to $>10^{7}$ K) to cold and even molecular gas [31,32].  Even though CGM densities are low, these regions are physically very large (e.g., [33]) and harbor an important fraction of the baryonic mass in galaxies [34].   To make progress toward understanding the full baryon cycle, a large space telescope with greater collecting area than \textsl{HST} is necessary, but high spectral resolution is also crucial.  Existing COS and STIS observations effectively demonstrate the importance of high spectral resolution for CGM studies, and in this section we present several examples that show the limitations and systematic errors that result from inadequate spectral resolution.

\vspace{5mm}
\begin{figure*}
\includegraphics[width=\textwidth]{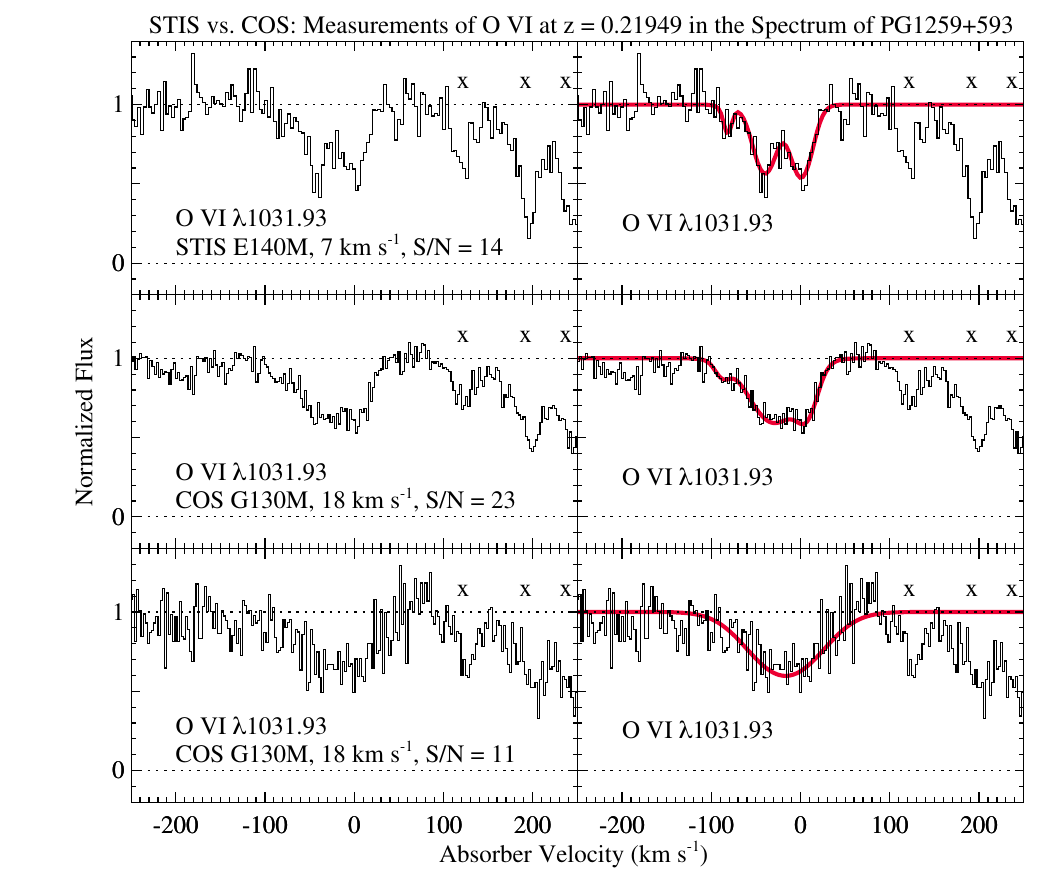}
\caption{\textsl{Left panel:} Comparison of STIS and COS observations of the absorption profile of the O~VI 1031.92 \AA\ line at $z_{abs} = 0.21949$ in the spectrum of the quasar PG1259+593. The upper-left panel shows the STIS E140M spectrum studied by Richter et al. (2004) and Tripp et al. (2008).  This spectrum has 7 km s$^{-1}$ spectral resolution and signal-to-noise (S/N) = 14 per resolution element.  The middle-left panel shows the COS G130M spectrum of the same profile obtained from the \textsl{HST} archive; this COS spectrum has 18 km s$^{-1}$ resolution and S/N = 23.  To illustrate the impact of lower S/N, the bottom-left panel shows a single COS G130M exposure of the same feature with S/N = 11.    \textsl{Right panels:} The same profiles shown in the corresponding left panels with Voigt-profile fits overlaid with the smooth red curves. In all panels, unrelated absorption features from the Milky Way ISM are marked with an `x'. \label{fig_pg1259}}
\end{figure*}

First, in Figure~\ref{fig_pg1259} we compare STIS E140M (7 km~s$^{-1}$ resolution) and COS G130M (18 km~s$^{-1}$ resolution) spectra of the QSO PG1259+593, including a high-signal-to-noise (S/N) version of the COS spectrum obtained by coadding all of the exposures available from the \textsl{HST} archive, and a low-S/N version obtained by extracting the spectrum from a single COS exposure. We compare measurements obtained from the high- and low-S/N versions below.  Looking first at the STIS E140M data, we see that this sight line has a redshifted O~VI absorption at $z_{abs} = 0.21949$ that clearly shows two well-detected components in the STIS O~VI 1031.92 profile (also evident in the O~VI 1037.62 \AA\ line not shown in Figure~\ref{fig_pg1259}, see [35], and a very weak third component at $v \approx -80$ km~s$^{-1}$ is also detected [36]. But in the 18 km~s$^{-1}$ resolution COS data, the two main components are not obvious despite good S/N of 23 per resolution element. Careful inspection of the COS profile reveals a flat bottom and asymmetric profile edges (as well as the weak third component), so one might be motivated to fit three components, although many researchers would probably choose only two components.  

If we do fit the high-S/N COS data in Figure~\ref{fig_pg1259} with three components, we obtain the results summarized in Table~\ref{cgm_fit_comparison}, where the STIS fitting results are also summarized.  Comparing the STIS vs. high-S/N COS results component by component, we see that the measurements are in rough agreement, but in several cases the parameters disagree by factors of $\approx$ 2, and these discrepancies are $2\sigma - 3\sigma$ larger than expected based on the formal model error bars.  It is evident that the STIS E140H spectra are better.  For example, inspection shows that the two main (stronger) components 2 and 3 have similar line widths and depths (see upper-left panel of Fig.~\ref{fig_pg1259}), so the Doppler parameters and column densities should be similar. This is indeed what is found in the STIS model fit.  But in the COS fit, component 2 has a very different velocity and line width than for the STIS data, and component 3 has $b_{3} = 14\pm4$ and log $N_{3} = 13.42\pm0.21$, a significant departure from the component properties evident in the higher-resolution spectra.  

Systematic errors at the factor of 2 level can lead to unfortunately specious conclusions.  For example, as discussed in Section~\ref{sec_mean_thermal}, line widths can be used to infer the temperature and non-thermal broadening (e.g., due to turbulence) of the gas.  Consider component 2 in Table~\ref{cgm_fit_comparison}.  If this line were predominantly thermally broadened, the high-S/N COS line width would imply a temperature of $\approx 1.2 \times 10^{6}$ K, which is close to the virial temperature of the halo of a Milky Way-like galaxy.  However, the STIS component-2 line width implies a temperature that is a factor of 4 lower, $T \approx 2.8 \times 10^{5}$ K.  The STIS temperature is closer to the range where O~VI is expected to be found in collisional ionization equilibrium (CIE) and is close to the peak in the cooling curve where the gas can cool rapidly.  The systematic error introduced in the COS measurement can lead to seriously misleading conclusions about the gas physics.  Similar systematic problems can occur when comparing absorption lines of different species in order to infer the non-thermal/turbulent broadening: errors in velocity centroids and line widths can lead to incorrect inferences and erroneously assigning different ions to the same gas phase.  Similarly, factors of $\approx$2 errors in column densities can cause incorrect conclusions in modeling the ionization mechanisms and physical conditions of the gas and plasma.  

%\begin{center}
\begin{table}
\small
\caption{Comparison of STIS vs. COS Measurements of PG1259+593 O~VI Absorption} 
\label{cgm_fit_comparison}
\begin{center}
\begin{tabular}{clll}
%\label{cgm_fit_comparison}
%\multicolumn{4}{c}{Table 1: Comparison of STIS vs. COS Measurements of PG1259+593 O~VI Absorption} \\ \hline \hline
 \  &  \                   & \                     & \\ \hline
 \ & STIS E140M & COS G160M & COS G160M \\
 PARAMETER$^{a}$ & (S/N = 14)     & (S/N = 23)     & (S/N = 11) \\ \hline
$v_{1}$ (km s$^{-1}$) & $-82\pm2$ & $-88\pm 3$ & \\
$v_{2}$ (km s$^{-1}$) & $-40\pm2$ & $-30\pm 4$ & $-17\pm 3$  \\
$v_{3}$ (km s$^{-1}$) & $1\pm1$    & $7\pm 2$ & \\
$b_{1}$ (km s$^{-1}$) & $6^{+6}_{-3}$ & 10$^{+8}_{-5}$ & \\
$b_{2}$ (km s$^{-1}$) & $17\pm3$ & 35$\pm$6 & $53\pm 4$ \\
$b_{3}$ (km s$^{-1}$) & $15\pm2$ & $14\pm4$ & \\ 
log $N_{1}$ & 12.85$\pm$0.16 & 12.79$\pm$0.21 & \\
log $N_{2}$ & 13.69$\pm$0.05 & 13.95$\pm$0.08 & $14.13\pm 0.03$  \\
log $N_{3}$ & 13.70$\pm$0.05 & 13.42$\pm$0.21 & \\ \hline
 \end{tabular}
 \end{center}
 
 \noindent {\footnotesize $^{a}$ Each component of a Voigt-profile fit is characterized by its velocity centroid ($v$), its line width expressed as a $b-$ value, and its column density $N$. Here the STIS and higher-S/N COS data have been fitted with three components, labeled with subscripts $1 - 3$, but the lower-S/N COS data only justify a single-compoent fit (see Figure~\ref{fig_pg1259}), so only one value is listed for each parameter.}
 \end{table}
 
The previous paragraph compared COS spectra with relatively good S/N = 23 to the STIS data. However, the vast majority of COS data in the \textsl{HST} archive has significantly lower S/N; most of the COS data have S/N closer to 10, similar to the low-S/N PG1259+593 data and measurements in Figure~\ref{fig_pg1259} and Table~\ref{cgm_fit_comparison}. Low 
S/N data can greatly exacerbate systematic errors in low resolution data, as we can see from Table~\ref{cgm_fit_comparison}.  At lower S/N, it is impossible to recognize the three components in the O~VI profile, and only a single component can be reasonably fitted.  With a single-component model, the inferred velocity centroid and line width are very far from the values found with higher resolution (see Table~\ref{cgm_fit_comparison}), and in this case only the total column density can be reliably measured. 
 
 \vspace{5mm}
 \begin{figure*}
\includegraphics[width=\textwidth]{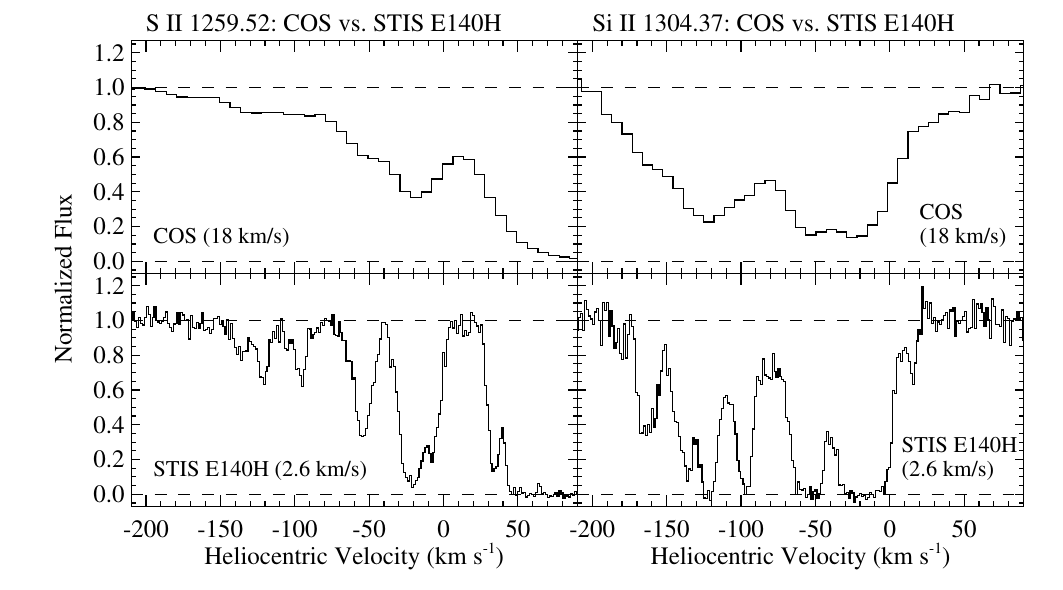}
\includegraphics[width=\textwidth]{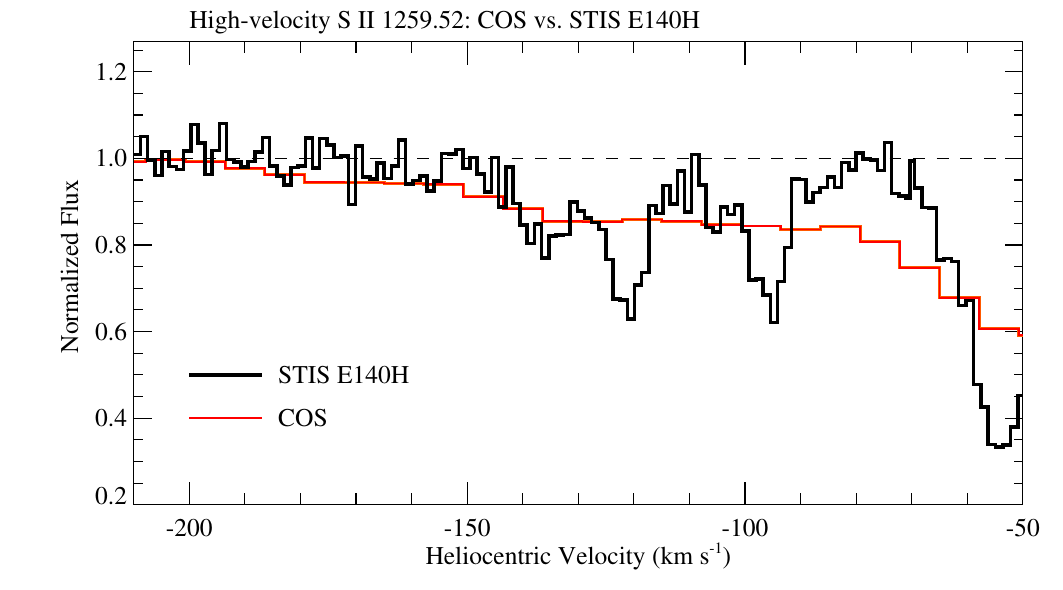}
\caption{\textsl{Upper panels:}  Comparison of very high signal-to-noise COS spectra (spectral resolution = 18 km s$^{-1}$) of MRK817 (upper left and upper right) to STIS E140H spectra (resolution = 2.6 km s$^{-1}$) of the same target and same lines.  The panels on the left compare the observed profiles of the Milky Way S~II 1259.52 \AA\ transition vs. Heliocentric velocity, and the right panels compare the Milky Way Si~II 1304.37 \AA\ transition.  The COS data have S/N $\gg$ 100 per resolution element. \textsl{Lower panel:} Zoomed in comparison of the S~II profiles in the high-velocity absorption components at $v_{\odot} < -100$ km s$^{-1}$.  At higher resolution, multiple high-velocity components are revealed.  \label{fig_mrk817}}
\end{figure*}

But is the 7 km s$^{-1}$ resolution provided by the STIS E140M spectrograph sufficient for all CGM science?  The answer is probably not.  Consider the Si~IV ion.  In CIE, this ion peaks in abundance at $T \approx 9\times 10^{4}$ K (see Figure 1 in [37].  At this temperature, the Si~IV $b-$value is 7.3 km s$^{-1}$, which is marginally resolved with E140M resolution.  However, many papers have argued that highly ionized CGM species such as C~IV, Si~IV, and O~VI are best explained by non-equilibrium ionization (e.g., [38,39] and references therein), so these species may be at much lower temperatures (see example non-equilibrium models Figure 1 in [37] where their lines are not adequately resolved at 7 km s$^{-1}$ resolution.  Moreover, O~VI absorption systems often have affiliated lower ionization stages that are kinematically correlated with the highly ionized gas; for example, the O~VI absorber shown in Figure~\ref{fig_pg1259} has affiliated C~III, O~III, O~IV, and Si~III [35].  These lower ionization species must exist in cooler gas phases (see [40]) and thus will have narrower line widths that cannot be adequately resolved at 7 km s$^{-1}$.  This has recently been demonstrated by [38], which compared STIS E140H (2.6 km s$^{-1}$) and STIS E140M (7 km s$^{-1}$) observations of a Milky Way high-velocity cloud (HVC) showing that lines that appear to be single components in the E140M data resolve into \textbf{five} discrete components in the E140H spectra.  This has important science consequences, as discussed in [38].  

Figure~\ref{fig_mrk817}  shows similar but more recent STIS E140H observations of absorption from high-velocity cloud Complex C, compared to very high-S/N COS observations of the same background AGN.  We see that in the S~II and Si~II lines, the seemingly simple component structure apparent in the COS spectra resolves into many intricate components.  In the velocity range of Complex C, the COS data suggest that there are three high-velocity components, but the STIS show that there are at least six components in the HVC.  From Figure~\ref{fig_mrk817} it is clear that the line widths, component centroids, and column densities of the individual components cannot be accurately measured, even in COS spectra with extraordinarily high S/N.  

Figure~\ref{fig_mrk817} shows another important source of systematic error in COS data. In the COS spectra, the cores of many of the features are well above the zero-flux level, and this would often be interpreted as indicating that the lines are not saturated.  However, when the same lines are measured with STIS E140H, we see that several of those features are, in fact, saturated and even strongly saturated.  The lower spectral resolution of COS, combined with wings in its line-spread function, causes the cores of saturated lines to be filled in.  Unresolved saturation can be recognized in low-resolution data when multiple transitions with differing line strengths can be compared [41,42], but if there is only a single transition available (e.g., for Al~II and Si~III, only one transition can be observed), observers should be extra cautious about how saturation could systematically degrade the measurements.

\textbf{Science Objective} We see from figure~\ref{fig_mrk817}  that the STIS E140H spectrograph provides excellent high-resolution data. Unfortunately, due to the limited collecting area of \textsl{HST} and the limited sensitivity of its old instruments, STIS in E140H mode can only observe the very brightest extragalactic targets, and the number of CGM studies that can be done with this instrument is limited.  Therefore it is clear that \textbf{high spectral resolution with resolving power $>$100,000, coupled with a telescope with larger collecting area and better instrument throughput than the HST spectrographs, is a fundamental requirement for future UV studies of the CGM and the baryon cycle of galaxies.}  A central objective of HWO is to study the CGM \underline{with statistically useful large samples}.  However, in order for those samples to deliver reliable measurements, the HWO spectrograph must have spectral resolution comparable to the STIS E140H resolution.

\section{Conclusions}

Narrow absorption features and closely spaced velocity structures may be hidden in moderate resolution (R=20,000--50,000) spectra, or if only partially resolved yield inaccurate column densities and abundances. One never really knows what new information a spectrum with resolution of R=100,000 (3~km~s$^{-1}$) or more would contain without obtaining the spectrum. For example, thermal broadening of cold star-forming gas at 100~K is about 1~km/s, and both flows and turbulence could have velocities similar to this value. In this paper we have shown that high-resolution spectra in the ultraviolet are required for understanding both inner and outer regions of the baryon cycle.

\begin{quotation}
When observing hot plasmas, low resolution is the score.

Observing cold and warm gas requires much more.

Not R=20,000 or even 50,000 as before.

Quoth the raven, ``Nevermore" 
\end{quotation}

(With apologies to Edgar Allen Poe's``The Raven".)

\end{spacing}

\noindent {\bf Code and Data Availability}\\

Data sharing is not applicable to this article as no new data were created or analyzed.\\

\noindent {\bf Disclosures}\\

The authors declare that there are no financial interests, commercial affiliations, or other potential conflicts of interest that could have influenced the objectivity of this research or the writing of this paper. The data shown in the figures are available from the authors.

\noindent {\bf Acknowledgments}\\

\noindent Data cited in this paper are based on observations with the NASA/ESA Hubble Space Telescope
obtained the Mikulski Archive for Space Telescopes at the Space Telescope Science
Institute, which is operated by the Association of Universities for
Research in Astronomy, Incorporated, under NASA contract NAS5-
26555. Support for Program number HST-GO-17227.001-A was
provided through a grant from the STScI under NASA contract NAS5-26555.
We acknowledge support from the NASA Outer Heliosphere Guest Investigators Program to Wesleyan University and the University of Colorado for grant 80NSSC20K0785.\\
 
\noindent {\bf References}\\
%\bibliography{report}   % bibliography data in report.bib

1. S. Redfield and J. L. Linsky, "The Structure of the Local Interstellar Medium. I. High-Resolution Observations of Fe II, Mg II, and Ca II toward Stars within 100 Parsecs" {\em Astrophys. J. Supl. Ser.} {\bf 139(2)}, 439--465 (2002).\\

2. R. M. Crutcher ``The local interstellar medium" {\em Astrophys. J.} {\bf 254}, 82--87 (1982).\\

3. R. Lallement and P. Bertin, ``Northern-hemisphere observations of nearby interstellar gas : possible detection of the local cloud" {\em Astron. Astrophys.} {\bf  266}, 479--485 (1992).\\

4. S. Redfield and J. L. Linsky, ``The Structure of the Local Interstellar Medium. IV. Dynamics, Morphology, Physical Properties, and Implications of Cloud-Cloud Interactions"  {\em Astrophys. J.} {\bf 673}, 283--324 (2008).\\ 

5. C. Gry and E. B. Jenkins, ``The interstellar cloud surrounding the Sun: a new perspective"  {\em Astron. Astrophys.} 
{\bf 567}, 58--75 (2014).\\

6. C. Malamut, S. Redfield, J. L. Linsky, B. E. Wood,  T. R. Ayres, ``The Structure of the Local Interstellar Medium. VI. New Mg II, Fe II, and Mn II Observations toward Stars within 100 pc"  {\em Astrophys. J.} {\bf 787}, 75 (2014).\\

7. S. Redfield and J. L. Linsky, ``The Structure of the Local Interstellar Medium. II. Observations of D I, C II, N I, O I, Al II, and Si II toward Stars within 100 Parsecs"  {\em Astrophys. J.}  {\bf 602}, 776 (2004).\\

8. J. L. Linsky and S. Redfield, ``Anomalous Neutral Hydrogen Column Densities in Local Interstellar Medium Clouds"
 {\em Astrophys. J.} to appear (2025).\\

9. B. E. Wood, J. L. Linsky,  G. H\'ebrard, et al., ``Two New Low Galactic D/H Measurements from the Far Ultraviolet Spectroscopic Explorer" {\em Astrophys. J.} {\bf 609}, 838--853 (2004).\\

10. J. L. Linsky, S. Redfield, D. Ryder, A. Chasan-Taber, ``Inhomogeneity within Local Interstellar Clouds" {\em Astron. J.} {\bf 164}, 106--130 (2022).\\

11. J. M. Shull and S. R. Kulkarni, ``Interstellar Bow Shocks around Fast Stars Passing through the Local Interstellar Medium" {\em Astrophys. J.} {\bf 951}, 35--40 (2023).\\

12. P. Swaczyna, N. A. Schwadron,  E. M\"ob ius, et al., ``Mixing Interstellar Clouds Surrounding the Sun" {\em Astrophys. J.}  {\bf 937}, L32 (2022).\\ 

13. C. Zucker, S. Redfield, S. Starecheski, R. Konietzka,  J. L. Linsky, ``The Origin of the Cluster of Local Interstellar Clouds"  {\em Astrophys. J.} submitted (2024).\\

14. E. B. Jenkins, and T. M. Tripp, ``The Distribution of Thermal Pressures in the Diffuse, Cold Neutral Medium of Our Galaxy. II. An Expanded Survey of Interstellar C I Fine-structure Excitations"  {\em Astrophys. J.}  {\bf 734}, 65--97 (2011).\\

15. E. B. Jenkins, and T. M. Tripp, ``Thermal Pressures in the Interstellar Medium away from Stellar Environments" {\em Astrophys. J.} {\bf 916}, 17 (2021).\\

16. E. B. Jenkins, and T. M. Tripp, ``The Distribution of Thermal Pressures in the Interstellar Medium from a Survey of C I Fine-Structure Excitation " {\em Astrophys. J.} {\bf 137}, 297 (2001).\\

17. J. E. Owen and A. P. Jackson, ``Planetary evaporation by UV \& X-ray radiation: basic hydrodynamics" {\em Mon. Notices Roy. Astron. Soc.} {\bf 425}, 2931--2947 (2012).\\ 

18. H. Lammer,  A. St\"okl, N. V. Erkaev, et al., ``Origin and loss of nebula-captured hydrogen envelopes from `sub'- to `super-Earths' in the habitable zone of Sun-like stars"  {\em Mon. Notices Roy. Astron. Soc.} {\bf 439}, 3225 (2014).\\ 

19. G. Van Looveren, M. G\"udel, S. Boro Saikia, K. Kislyakova, ``Airy worlds or barren rocks? On the survivability of secondary atmospheres around the TRAPPIST-1 planets" {\em Astron. Astrophys.} {\bf 683}, A153--168 (2024).\\

20. V. R\'eville, J.M. Jasinski, M. Velli, et al., `` Magnetized Winds of M-type Stars and Star–Planet Magnetic Interactions: Uncertainties and Modeling Strategy" {\em Astrophys. J.} {\bf 976}, 65--77 (2024).\\

21. J. L. Linsky,  "Host Stars and their Effects on Exoplanet Atmospheres (Second Edition)". Springer Nature (2025).\\ 

22. K. France,  N. Arulanantham,  E. Maloney,  et al., `` The Radial Distribution and Excitation of H2 around Young Stars in the HST-ULLYSES Survey"  {\em Astron. J.} {\bf 166}, 67--81 (2023).\\ 

23. D. J. Teal, E. M.-R. Kempton, S. Bastelberger, A. Youngblood, G. Arney, ``Effects of UV Stellar Spectral Uncertainty on the Chemistry of Terrestrial Atmospheres" {\em Astrophys. J.} {\bf 927}, 90--108 (2022).\\

24. D. J. Wilson, C. S. Froning, G. M. Duvvuri, et al., ``The Mega-MUSCLES Treasury Survey: X-Ray to Infrared Spectral Energy Distributions of a Representative Sample of M Dwarfs"  {\em Astrophys. J.} {\bf 978}, 85--105 (2025).\\

25. R. O. P. Loyd, K. France,  A. Youngblood, et al., ``The MUSCLES Treasury Survey. III. X-Ray to Infrared Spectra of 11 M and K Stars Hosting Planets" {\em Astrophys. J.} {\bf 824}, 102--121 (2016)\\

26. H. Yang, J. L. Linsky, K. France, ``HST/COS Spectra of DF Tau and V4046 Sgr: First Detection of Molecular Hydrogen Absorption Against the Ly$\alpha$ Emission Line" {\em Astrophys. J. Lett.} {\bf  730}, L10-15 (2011).\\

27. A. Brennan,  M. Luca, M. Sebastian, et al.,  ``Low CI/CO abundance ratio revealed by HST UV spectroscopy of CO-rich debris discs" {\em Mon. Notices Roy. Astron. Soc.} {\bf 531}, 4482 (2024).\\

28. LUVOIR STDT, “LUVOIR Final Report,” https://asd.gsfc.nasa.gov/luvoir/reports/ (2019).\\

29. K. France, G. J. Herczeg, M. McJunkin, S. V. Penton,  `` CO/H$_2$ Abundance Ratio $\approx 10^{-4}$ in a Protoplanetary Disk" {\em Astrophys. J.} {\bf 794}, 160--172 (2014).\\

30. P. W. Cauley, K. France, G. J. Herczeg,  C. M. Johns-Krull, ``A CO-to-H$_2$ Ratio of $\approx10^{-5}$ toward the Herbig Ae Star HK Ori" {\em Astron. J.} {\bf 161}, 217 (2021).\\ 

31. J. Tumlinson, M. S. Peeples,  J. K. Werk,  ``The Circumgalactic Medium" {\em Ann. Rev. Astron. Astrophys.} {\bf 55}, 389--432 (2017).\\

32. M. Donahue and  G. M.Voit, ``Baryon cycles in the biggest galaxies" {\em Phys. Rep.} {\bf 973}, 1 (2022).\\

33. J. Tumlinson, C. Thom, J. K. Werk, et al., ``The Large, Oxygen-Rich Halos of Star-Forming Galaxies Are a Major Reservoir of Galactic Metals" {\em Science} {\bf 334}, 948 (2011).\\

34. J. K. Werk, J. X. Prochaska, J. Tumlinson, et al., ``The COS-Halos Survey: Physical Conditions and Baryonic Mass in the Low-redshift Circumgalactic Medium" {\em Astrophys. J.} {\bf 792}, 8 (2014).\\

35. P. Richter, B. D. Savage, T. M. Tripp, K. R.  Sembach,  ``FUSE and STIS Observations of the Warm-hot Intergalactic Medium toward PG 1259+593" {\em Astrophys. J. Supl. Ser.} {\bf 153}, 165 (2004).\\

36. T. M. Tripp, K. R. Sembach, D. V. Bowen, et al., ``A High-Resolution Survey of Low-Redshift QSO Absorption Lines: Statistics and Physical Conditions of O VI Absorbers" {\em Astrophys. J. Supl. Ser.}  {\bf 177}, 39--102 (2008).\\

37. N. Lehner, and J. C. Howk, `` A Reservoir of Ionized Gas in the Galactic Halo to Sustain Star Formation in the Milky Way" {\em Science} {\bf 334}, 955 (2011).\\

38. T. M. Tripp,  ``The high-velocity clouds above the disc of the outer Milky Way: misty precipitating gas in a region roiled by stellar streams" {\em Mon. Notices Roy. Astron. Soc.} {\bf 511}, 1714 (2022).\\

39. S. Kumar and  H.-W. Chen, ``Non-equilibrium ionization in the multiphase circumgalactic medium -- impact on quasar absorption-line analyses" arXiv:2501.13170 (2025).\\

40. Haislmaier, T. M. Tripp, N. Katz, et al., ``The COS Absorption Survey of Baryon Harbors: unveiling the physical conditions of circumgalactic gas through multiphase Bayesian ionization modelling" {\em Mon. Notices Roy. Astron. Soc.} {\bf 502}, 4993 (2021).\\

41. B. D. Savage and  R. R. Sembach, ``The Analysis of Apparent Optical Depth Profiles for Interstellar Absorption Lines"
 {\em Astrophys. J.} {\bf  379}, 245 (1991).\\

42. E. B. Jenkins,  ``A Procedure for Correcting the Apparent Optical Depths of Moderately Saturated Interstellar Absorption Lines" {\em Astrophys. J.}  {\bf 471}, 292 (1996).\\

\vspace{2ex}\noindent\textbf{Jeffrey Linsky} is Professor Emeritus in the Department of Astrophysical and Planetary Sciences at the University of Colorado in Boulder and a Fellow (now retired) of JILA, a joint institute of NIST and the University of Colorado. He received his BS degree in physics from MIT in 1963 and his PhD in astronomy from Harvard University in 1968. He is the author of two books, more than 450 journal papers, and many book chapters. His current research interests include the interstellar medium, the interactions of stars and their exoplanets, and stellar winds. He is a Fellow of the AAS and a member of the IAU.\\

\end{document}